\documentclass{ifacconf}

\usepackage{graphicx}      
\usepackage{natbib}        
\usepackage{amssymb}      
\usepackage{amsmath}      
\usepackage{xcolor}       
\usepackage{float}        
\usepackage{tikz}         
\usepackage{url}          
\usetikzlibrary{positioning,calc} 
\usepackage{eso-pic}
\AddToShipoutPictureBG*{%
\AtPageUpperLeft{%
\setlength\unitlength{1in}%
\hspace*{\dimexpr0.5\paperwidth\relax}
\makebox(0,-1.75)[c]{
\begin{tabular}{c c}
Marzhan M. Baubekova \emph{et al.},
Supervisor synthesis for multilevel DES with local buses, \\
To appear in
{\em 23rd IFAC World Congress},
Busan, South Korea, 2026, 
uploaded to ArXiv \today \\
\end{tabular}}}}
\tikzset{
  connector/.style={thick},%
  bus/.style={
    draw,
    rectangle split,
    rectangle split horizontal,
    rectangle split parts=2,
    minimum width=0.2cm,
    minimum height=0.2cm,
    align=center
  }
}
\usepackage{algorithm}
\usepackage{algcompatible}
\begin{document}
\begin{frontmatter}

\title{Supervisory control synthesis for multilevel DES with local buses\thanksref{footnoteinfo}} 

\thanks[footnoteinfo]{This work is supported by Rijkswaterstaat, the executive agency of the Dutch Ministry of Water Management and Infrastructure.}

\author[First]{Marzhan M. Baubekova} 
\author[First]{Martijn A. Goorden} 
\author[First]{Michel A. Reniers}
\author[First]{Joanna M. v.d. Mortel-Fronczak} 
\author[First]{Jacobus E. Rooda} 
\author[First]{Wan J. Fokkink}
\address[First]{Department of Mechanical Engineering, Eindhoven University of Technology, The Netherlands.(e-mail: m.baubekova@tue.nl).}

\begin{abstract}                
In multilevel supervisor synthesis, dependency structure matrix techniques can be used to transform the models of plants and requirements into a tree-structured hierarchical decomposition of the synthesis problem and thus efficiently synthesize local supervisors. A bus component, which has many dependencies across a system, tends to lead to an undesirable clustering of many components in one synthesis subproblem. Prior work showed how to recognize and properly treat a global bus structure. In this paper, we leverage this work from global to local bus structures through a novel multilevel discrete-event system (MLDES) architecture. Specifically, the hierarchical system decomposition is revisited by allowing bus detection not only on the top level but at each level of the system hierarchy. Given this architecture, an algorithm is introduced that constructs a tree-structured MLDES. A case study on a production line shows the effectiveness of the proposed method through significantly improved synthesis performance, measured by the sum of the controlled state-space sizes of the local supervisors.

\end{abstract}


\end{frontmatter}

\section{Introduction}
The supervisory control theory (SCT), initially proposed in \cite{ramadge-1982}, provides a formal framework to synthesize a supervisor. The key advantage of SCT is that it clearly separates the concept of open-loop dynamics (the plant) and feedback control, shifting the focus in the controller design from \textit{how} the system should operate to \textit{what} the system should do. By providing a discrete-event model of the uncontrolled system and control requirements (the \textit{what} part), a correct-by-construction supervisor can be automatically constructed (the \textit{how} part). However, for large-scale discrete-event systems (DES) supervisor synthesis is limited by high computational memory demand \citep{wonham-2018}. 
To address this issue, various supervisory control architectures have been proposed in the literature, all aimed at dividing the overall synthesis problem into smaller subproblems. 

\cite{ramadge-1989} and \cite{deQueiroz-2000} suggest modular synthesis to reduce the complexity of DES models, by constructing a separate supervisor for each individual requirement. \cite{ramadge-1989} and \cite{yoo-2002} extend the idea to the decentralized architecture, by decomposing the synthesis problem according to the plant structure, so that local supervisors are constructed, each observing and controlling only a subset of the plant. \cite{zhong-1990} and \cite{leduc-2009} combine decomposition- and abstraction-based approaches in a hierarchical architecture, organizing the plant and the supervisor into layers where higher-level controllers coordinate or restrict the behavior of lower-level ones. \cite{Komenda-2013, Komenda-wodes16} propose a multilevel approach to modular supervisor synthesis relying on vertical and horizontal decomposition of the system, structuring supervisors in hierarchical layers without explicit abstraction between them. Structuring large-scale systems so that these methods can be applied properly is a challenging task that requires additional information about the system's or controller's structure.  

This paper focuses on multilevel synthesis as it requires a tree-structured system, which reflects the natural way engineers decompose and approach complex systems. The monolithic problem is decomposed into subproblems, such that for each subsystem a supervisor is synthesized based on requirements for only that subsystem. The decomposition can be guided by a Dependency Structure Matrix (DSM) as proposed in \cite{Goorden-2020}, which reveals the structure of the system based on the dependencies determined by the requirements between the components. The current findings using DSM-based techniques show a significant reduction in the sum of the controlled state-space sizes of multilevel supervisors compared to the controlled state-space size of a monolithic supervisor. An automatically derived MLDES using DSM-based techniques is also more beneficial compared to the manually constructed MLDES demonstrated by \cite{Reijnen-ccta18}. In \cite{Goorden-ecc19}, bus structure is incorporated in the previous MLDES architecture. Inspired from computer science, where a bus is a communication system that transfers data between components within a computer, in control architecture a bus includes the components that have dependencies with many components across the system. The experimental results on the benchmark models obtained by \cite{Goorden-ecc19} revealed that having a global bus architecture can be beneficial, but only for systems that have clear bus components. For the rest of the cases, the MLDES without a global bus produced better results.


This paper contributes by proposing a new concept to the DSM-based clustering technique and a systematic procedure for the construction of MLDES for multilevel synthesis of \cite{Komenda-wodes16}. Thinking of a single global bus limits its applicability to systems where the bus components can be detected at lower levels of decomposition, such as those characterized by sequential control. In a DES exhibiting sequential control, the system behavior is driven by a series of events that occur in a specific order, creating a chain of dependencies. To address this limitation, we extend the MLDES architecture to allow the detection of buses at arbitrary hierarchical depths. In this paper, we adopt the clustering algorithm of \cite{wilschut-2017} to identify buses and clusters, but the proposed method is agnostic to the specific clustering approach. Subsequently, local buses are distributed over the MLDES being placed as deep as possible in the clustering hierarchy. By doing so, we can achieve multilevel synthesis subproblems of more balanced size, distributing the computational effort more evenly. The benefit in terms of the reduced controlled state-space size of the multilevel supervisors is demonstrated in two use cases: a production cell \citep{feng-2009} and a production line \citep{Reijnen-ccta18}. 

The rest of the paper is structured as follows. Section \ref{s:background} provides the concepts and notions of supervisor synthesis, multilevel synthesis, and DSMs. A new definition of MLDES architecture with local buses using the production cell as an example is given in Section \ref{s:localbus}. To transform it into a MLDES, a supporting algorithm is explained in Section \ref{s:algo}. Section \ref{s:example} presents a case study on the production line system. Finally, Section \ref{s:conclusion} concludes the paper and provides further research suggestions.

\section{Preliminaries}\label{s:background}

\subsection{Supervisory control}
The supervisory control synthesis problem is formalized as the design of a supervisor which, on observing the events generated by the plant, would dynamically disable a suitable subset of controllable events to guarantee that the resulting controlled behavior adheres the specified behavior. Supervisory control theory (SCT)  provides a framework to synthesize a supervisor for discrete-event systems (DES) given models of the uncontrolled system and of the control requirements \citep{cassandras-2021,ramadge-1982,ramadge-1987,ramadge-1989, wonham-2019}. Monolithic supervisory control synthesis derives a single supervisor based on a single plant model and a single requirement model \cite{ramadge-1987}. The synthesized supervisor is correct-by-construction and satisfies the following control objectives.
\begin{itemize}
    \item Safety: it adheres to all modelled requirements.
    \item Controllability: it restricts only controllable events.
    \item Nonblockingness: it can always continue to a marked state.
    \item Maximal permissiveness: it does not restrict more than necessary.
\end{itemize}
Originally, SCT could only handle the plant and the requirements modeled using finite automata (FA). Subsequent developments introduced more advanced, yet equally expressive, state-based requirement models in \cite{ma-2006,markovski-2010}. Later, \cite{Ouedraogo-2011} extended the theory to handle extended finite automata (EFA) that incorporate variables. 


For large-scale systems, plants and requirements are modeled as a collection of EFAs representing a \textit{composed system}, see \cite{mohajerani-2016}. 
Any composed system can be transformed into a product system by restructuring its EFAs so that each pair of EFAs has disjoint event and variable sets \citep{Goorden-2020}. In product system representation, no synchronization can occur between EFAs. The finest decomposition of a composed system with the maximum possible number of asynchronous EFAs is called the \textit{most refined product system}, see \cite{Goorden-2020}. 


\subsection{Multilevel supervisor synthesis}
A multilevel discrete-event system (MLDES) was introduced in \cite{Komenda-wodes16} and represents a system with a tree-based structure, $T$, where each tree node $n\in T$ has a set of children at the next lower level (except for the leaf node) and a unique parent at the next higher level (except for the top node). Each node in a tree is regarded as a subsystem consisting of subset of plant component models, $\{G_n\mid n\in T\}$, and corresponding requirements models, $\{K_n\mid n\in T\}$. 
As suggested in \cite{Komenda-wodes16}, the set of supervisors can be constructed by synthesizing for each node a supervisor with monolithic supervisory control synthesis. Then, their synchronous composition results in a global supervisor, formally defined as $S = ||_{n \in T} S_n$. In \cite{Komenda-wodes16}, it was shown that the controlled system, $S||G$, where $G = ||_{n \in T} G_n$ is the global plant, satisfies safety and controllability. Additionally, nonblockingness and maximal permissiveness can be ensured for a set of prefix-closed control requirements \citep{Komenda-wodes16, cassandras-2021}. Otherwise, nonblockingness of the composed supervised system needs to be verified as it is done in modular supervisor synthesis. 

\subsection{Dependency structure matrices}
However, structuring MLDES is not so intuitive from the control perspective, as the dependencies between plant components are dictated by the control requirements. As it was proposed in \cite{Goorden-2020, Goorden-ecc19}, DSMs aid the structuring of large DESs. As a basis for construction of the MLDES, a multilevel clustering is proposed. \cite{Goorden-2020} defines the multilevel clustering as recursively \textit{partitioning} set $A$ into $\{A_1, . . .,A_s\}$ with each cell $A_i$ being again partitioned, continued until partitions with a single element are reached. 
\begin{defn}
     (Multilevel clustering \citep{Goorden-2020}). The set of all multilevel clusterings $C^m_A$ on a non-empty element set $A$ is inductively defined.
\begin{itemize}
    \item When $|A| = 1,$ $(A,A)\in C^m_A.$
    \item For all $i\in \{1,...,s\}$: $(A_i,M_i)\in C^m_{A_i}$ and all $A_i$ being pairwise disjoint, we define $(A,M) \in C^m_A$ with $A=\bigcup\limits^s_{i=1}A_i$ and $M=\{(A_i,M_i)\mid 1\leq i \leq s\}$.
\end{itemize}
\end{defn}

The next steps describe how to find a multilevel clustering $(I,M)$, where $I$ is the index set of system components and $M$ is the set of multilevel clusterings of its children, given the most refined product system $\{G_i \mid i\in I\}$ with $G = ||_{i\in I} G_i$, $I=\{1,2,..,g\}$, $g\in\mathbb{N}^+$ and the set of control requirements $\{K_j \mid j\in J\}$. 
First, the dependencies between the most refined product system $\{G_i\mid i \in I\}$ and the set of control requirement models $\{K_j\mid j\in J\}$ are recorded in a Domain Mapping Matrix (DMM), denoted by $\mathit{PR}$. This matrix $\mathit{PR}$ is constructed as a binary matrix in which $PR(i,j) = 1$ indicates that $G_i$ and $K_j$ share an event or a variable. From DMM $\mathit{PR}$ a DSM, denoted by $P$, is derived. This matrix $P$ is calculated as a multiplication of $\mathit{PR}$ by its transpose, and represents the dependency between plant components. Entry $P(a,b)$ corresponds to the number of requirements that use events or variables from both $G_a$ and $G_b$. The clustering algorithm of \cite{wilschut-2017} is applied to $\mathit{P}$, revealing a multilevel clustering structure through manipulations of its rows and columns. The algorithm requires the user to manually tune the following parameters: expansion coefficient $\alpha$, inflation coefficient $\beta$, and evaporation constant $\mu$.  Finally, an MLDES is created from the multilevel clustering, where the requirements are placed at the nodes containing its associated plants.


\cite{Goorden-ecc19} proposes a further advancement that includes a special treatment of components that have dependencies with many other components across the system, a so-called \textit{bus}. Similarly, using the clustering algorithm of \cite{wilschut-2017} by tweaking an additional parameter $\gamma$, the bus and non-bus components are identified and clustered separately. Subsequently, two MLDESs are constructed and merged together, where bus plant models are distributed over the combined MLDES.

When comparing the two MLDESs with and without bus structure based on the experiment involving benchmark models described by \cite{Goorden-ecc19}, the latter reduces the computational effort of the synthesis problem only for systems with clear bus structure, but for the production cell and production line examples characterized by sequential control, the results are less favorable. 
 
\section{Proposed extension to multilevel control architecture}\label{s:localbus}
To introduce the proposed extension to the MLDES architecture and compare it with the existing MLDES architectures without and with global bus, this section presents a simple illustrative example, being a production cell from \cite{feng-2009}. This is one of the benchmark models tested in \cite{Goorden-ecc19}, where MLDES without global bus outperformed MLDES with global bus.

\begin{figure}
    \centering
    \includegraphics[width=7.5cm]{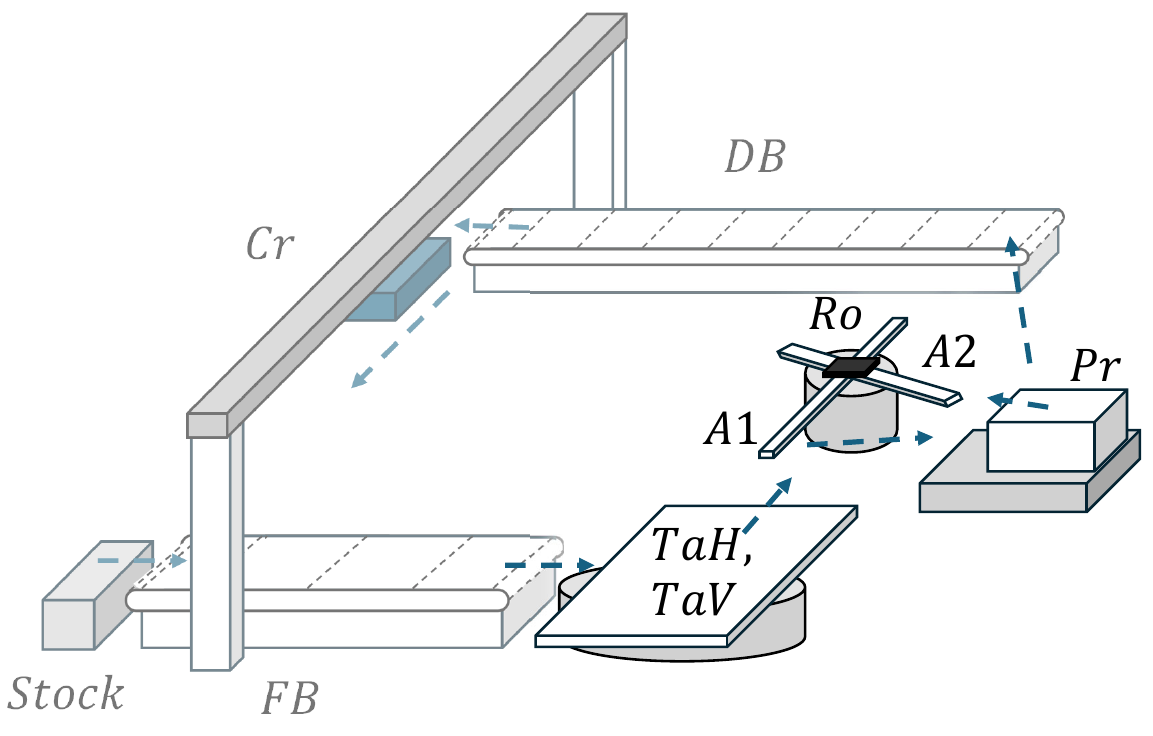}
    \caption{Overall system structure of a production cell with highlighted subsystems used as illustrative example. }
    \label{fig:prodcell}
\end{figure}

Fig.\ \ref{fig:prodcell} depicts an overall system layout with the highlighted subsystems, namely a robot ${Ro}$ with two arms ${A1}$ and ${A2}$ moving objects from the elevating rotary table ${TaV}$ and ${TaH}$ to the press ${Pr}$. 
 In the best performing MLDES architecture from \cite{Goorden-ecc19}, these subsystems are clustered together, and the multilevel synthesis subproblems arising from these clustered subsystems contribute the most to the overall computational effort. Thus, we consider them in isolation to clarify the distinction between multilevel clusterings and their effect on the multilevel synthesis performance. The simplified model consists of six components and 12 control requirements. The complete model, and other models used later in this paper, can be found in the accompanying GitHub repository \citep{github}. 

\begin{figure}
    \centering
    \begin{tikzpicture}[
      every node/.style={draw, rectangle, align=center, font=\tiny, minimum width=1cm, minimum height=0.5cm}
    ]
    
    \node (root) { $G:\{TaH,A1\}$\\
    $|R|: 1$ };
    \node[draw=black!0, right=0.01cm of root]{$css: 77$};
    \node (left) [below left=0.25cm and -0.5cm of root] {$G:TaH$};
    \node (right) [below right=0.25cm and -0.5cm of root] {$G:\{TaV,Ro,A1\}$\\
    $|R|: 2$};    
    \node[draw=black!0, right=0.01cm of right]{$css: 1,188$};
    
    \node (rightleft) [below left=0.25cm and -0.35cm of right] {$G:TaV$};
    \node (right2) [below right=0.25cm and -0.35cm of right] {$G:\{Pr,Ro,A1,A2\}$\\$|R|: 5$};
    \node[draw=black!0, right=0.01cm of right2]{$css: 16,071$};
    
    \node (right2left) [below left=0.25cm and -0.35cm of right2] {$G:\{Pr,Ro,A1\}$\\$|R|: 4$};
    \node[draw=black!0, right=0.01cm of right2left]{$css: 782$};
    \node (right3) [below right=0.25cm and -0.35cm of right2] {$G: A2$};
    
    \node (Pr) [below left=0.25cm and -0.35cm of right2left] { $G:Pr$};
    \node (Ro) [below=0.25cm of right2left] {$G:Ro$};
    \node (A1) [below right=0.25cm and -0.35cm of right2left] {$G:A1$};
    
    \newcommand{\orthconn}[2]{%
      \draw[connector]
        (#1.south) |- ($(#1.south)!.5!(#2.north)$) -| (#2.north);
    } 
    \orthconn {root}{left};
    \orthconn {root}{right};
    
    \orthconn {right}{rightleft};
    \orthconn {right}{right2};
    
    \orthconn {right2}{right2left};
    \orthconn {right2}{right3};
    
    \orthconn {right2left}{Pr};
    \orthconn {right2left}{Ro};
    \orthconn {right2left}{A1};
    
    \end{tikzpicture}
    \caption{The MLDES without bus.}
    \label{fig:prodcell-mldes1}
\end{figure}
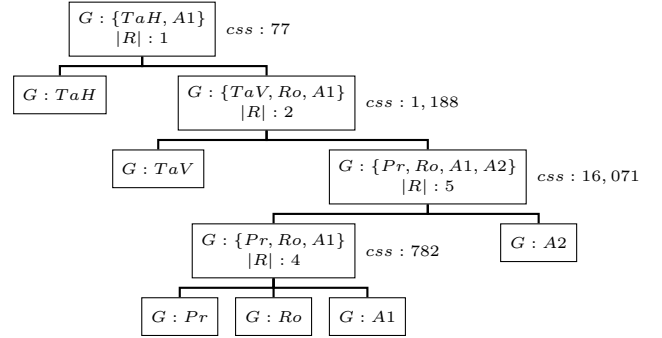

We apply three MLDES architectures on this example. First, in Fig.~\ref{fig:prodcell-mldes1} the MLDES is constructed from the multilevel clustering, where starting from the bottom: press, robot and arm $A1$ are clustered together, and with arm $A2$ they make the next level cluster, which is further clustered together with $TaV$, finally completing the system with $TaH$. The hierarchical structure of MLDES is identical to the structure of the multilevel clustering. Each node in the MLDES represents a synthesis subproblem, for each of which a monolithic supervisor can be synthesized. Synthesis of the leaf nodes that contain only plant components that have already been covered in the parent node can be omitted. The nodes for which the supervisor is synthesized include plant components $G$, the number of control requirements that reference them $|R|$, and the controlled state-space size of the resulting supervisor $css$. If we sum each node's $css$, this MLDES results in the total controlled state-space size of the multilevel supervisors of $18,118$.

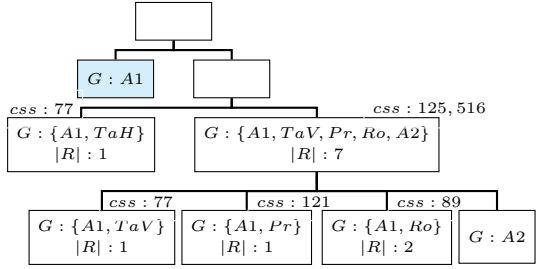
\begin{figure}
    \centering
    \begin{tikzpicture}[
      mainnode/.style={draw, rectangle,  align=center, font=\tiny, minimum width=1cm, minimum height=0.5cm},
      busnode/.style={draw, rectangle, fill=cyan!15, align=center, font=\tiny, minimum width=1cm, minimum height=0.5cm},
      level 1/.style={sibling distance=0.01mm, level distance=0.01mm},
      level 2/.style={sibling distance=0.01mm, level distance=0.01mm},
      node distance=0.001mm
    ]
    
    \node[mainnode] (root) {};
    
    \node[busnode] (A1) [below left=0.25cm and -0.25cm of root] {$G: A1$};
    \node[mainnode] (right) [below right=0.25cm and -0.25cm of root] {};
    
    \node[mainnode] (left2) [below left=0.25cm and 0.5cm of right] {$G:\{A1, TaH\}$\\ $|R|: 1$};
    \node[mainnode,draw=black!0, above left=-0.15cm and -1cm of left2]{$css: 77$};
    \node[mainnode] (right2) [below right=0.25cm and -1cm of right] {$G:\{A1, TaV, Pr, Ro, A2\}$\\ $|R|: 7$};
    \node[mainnode,draw=black!0, above right=-0.15cm and -1cm of right2]{$css: 125,516$};
    \node[mainnode] (TaV) [below left=0.5cm and 0.2
    5cm of right2] { $G:\{A1, TaV\}$\\$|R|: 1$};
    \node[mainnode,draw=black!0, above right=-0.15cm and -1cm of TaV]{$css: 77$};
    \node[mainnode] (Pr)  [below left=0.5cm and -1.55cm of right2] {$G:\{A1, Pr\}$\\$|R|: 1$};
    \node[mainnode,draw=black!0, above right=-0.15cm and -0.85cm of Pr]{$css: 121$};
    \node[mainnode] (Ro)  [below right=0.5cm and -1.55cm of right2] { $G:\{A1, Ro\}$\\$|R|: 2$};
    \node[mainnode,draw=black!0, above right=-0.15cm and -0.85cm of Ro]{$css: 89$};
    \node[mainnode] (A2)  [minimum height=0.7cm, below right=0.5cm and 0.25cm of right2] {$G:A2$};
    \newcommand{\orthconn}[2]{%
      \draw[connector]
        (#1.south) |- ($(#1.south)!.5!(#2.north)$) -| (#2.north);
    } 
    \orthconn{root}{A1}
    \orthconn{root}{right}
    \orthconn{right}{left2}
    \orthconn{right}{right2}
    \orthconn{right2}{TaV}
    \orthconn{right2}{Pr}
    \orthconn{right2}{Ro}
    \orthconn{right2}{A2}

    \end{tikzpicture}
    \caption{The MLDES with global bus (highlighted in blue).}
    \label{fig:prodcell-mldes2}
\end{figure}

One can observe that arm $A1$ is present in each non-leaf node of the MLDES given in Fig.\ \ref{fig:prodcell-mldes1}, implying that it has dependencies with many other components. Thus, we can identify $A1$ as a bus that results in the MLDES given in Fig.\ \ref{fig:prodcell-mldes2}. This MLDES is constructed from the multilevel clustering that at the top level has a global bus cluster and a non-bus cluster. The non-bus cluster is divided into two clusters, where the first cluster contains $TaH$ and the second cluster contains the remaining four components. However, the MLDES with the bus shows an increase in the state-space size of the multilevel supervisors to $125,880$. This can be explained by the fact that this system does not have a clear bus structure, since the robot also has many dependencies on other components and together they create a computationally large subproblem, including almost all components except $TaH$ and seven control requirements, which dominates the total $css$ ($125,516$). 

\begin{figure}
    \centering
    \begin{tikzpicture}[
      mainnode/.style={draw, rectangle,  align=center, font=\tiny, minimum width=1cm, minimum height=0.5cm},
      busnode/.style={draw, rectangle, fill=cyan!15, align=center, font=\tiny, minimum width=1cm, minimum height=0.5cm},
      level 1/.style={sibling distance=0.01mm, level distance=0.01mm},
      level 2/.style={sibling distance=0.01mm, level distance=0.01mm},
      node distance=0.001mm
    ]
    
    \node[mainnode] (root)[align=left] {};
    
    \node[busnode] (A1) [below left=0.25cm and -0.25cm of root] {$G:A1$};
    \node[mainnode] (right) [below right=0.25cm and -0.25cm of root] {};
    
    \node[mainnode] (left2) [below left=0.25cm and -0.5cm of right] {$G:\{A1, TaH\}$\\ $|R|: 1$};    
    \node[mainnode,draw=black!0, left=0.01cm of left2]{$css: 77$};
    \node[mainnode] (right2) [minimum height=0.7cm,below right=0.25cm and -0.25cm of right] {};
    \node[busnode] (left3) [below left=0.25cm and -0.25cm of right2] {$G:\{A1, Ro\}$\\ $|R|: 1$};    
    \node[mainnode,draw=black!0, left=0.01cm of left3]{$css: 89$};
    \node[mainnode] (right3) [minimum height=0.7cm,below right=0.25cm and 0cm of right2] {};
    \node[mainnode] (left4) [below left=0.25cm and -0.25cm of right3] {$G:\{A1, Ro, TaV\}$\\ $|R|: 2$};
    \node[mainnode,draw=black!0, left=0.01cm of left4]{$css: 1,188$};
    \node[mainnode] (right4) [below right=0.25cm and -0.5cm of right3] {$G:\{Ro, Pr, A2\}$\\$ |R|: 2$};
    \node[mainnode,draw=black!0, right=0.01cm of right4]{$css: 729$};
    \node[mainnode] (left5) [below left=0.5cm and -1cm of right4] {$G:\{A1, Ro, Pr\}$\\ $|R|: 2$};
    \node[mainnode,draw=black!0, above left=-0.15cm and -1cm  of left5]{$css: 945$};
    \node[mainnode] (right5) [below right=0.5cm and -1cm of right4] {$G:\{A1, Ro, A2\}$\\$ |R|: 3$};
    \node[mainnode,draw=black!0, above right=-0.15cm and -1cm of right5]{$css: 2,061$};
    \newcommand{\orthconn}[2]{%
      \draw[connector]
        (#1.south) |- ($(#1.south)!.5!(#2.north)$) -| (#2.north);
    } 
    \orthconn{root}{A1}
    \orthconn{root}{right}
    \orthconn{right}{left2}
    \orthconn{right}{right2}
    \orthconn{right2}{left3}
    \orthconn{right2}{right3}
    \orthconn{right3}{left4}
    \orthconn{right3}{right4}
    \orthconn{right4}{left5}
    \orthconn{right4}{right5}
    
    \end{tikzpicture}
    \caption{The MLDES with local buses (highlighted in blue).}
    \label{fig:prodcell-mldes3}
\end{figure}
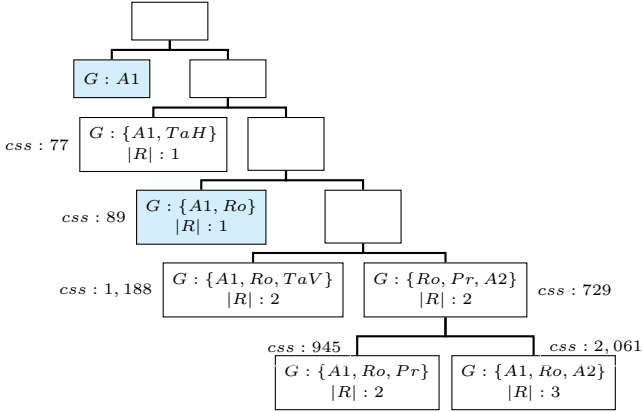

To reduce the synthesis problem we apply the bus detection algorithm to the previously discussed node. Fig.~\ref{fig:prodcell-mldes3} illustrates the resulting MLDES, with the restructured node, having robot $Ro$ identified as bus and the remaining components structured in two clusters, $TaV$ in the first cluster and $Pr$ with $A2$ in the second cluster. Then both the global and local buses are distributed over the tree and placed as low as possible. This propagation of the bus components to the lowest feasible level relieves the upper nodes from excessive synthesis effort, adding a marginally increased effort at lower levels. Consequently, the total number of states in the controlled MLDES decreased from 125,880 to 5,089, which also outperforms the MLDES without bus with a total $css$ of 18,118.

In the current clustering algorithm, the system components can be distributed into global bus and clusters, where each of them further can be decomposed into clusters as it was shown in the previous section.  In the proposed architecture, bus detection is also allowed within the clusters, so each cluster is treated as a separated system, which components can also be interpreted as a bus. Thus, we propose a new definition of a multilevel clustering. 


\begin{defn}\label{def:2}
(Multilevel clustering with local buses). The set of all multilevel clusterings $C_A^{mb}$ on a non-empty element set $A$ with index set $J=\{1,2,\ \ldots,\ k\}, k\in \mathbb{N}^{+}$ is inductively defined as follows.
\begin{itemize}
    \item When $|A| = 1$, $(A,\varnothing, A,R)\in C_A^{mb}$ or $(A,A,\varnothing,R)\in C_A^{mb}$ depending on being bus or non-bus cluster with \\$R\subseteq J$ denoting the index set of control requirements.
    \item For all $i \in \{1,...,s\}$ with $(A_i,B_i,M_i,R_i)\in C_{A_i}^{mb}$ and all $A_i$ being pairwise disjoint, we define $(A,B, M, R) \in C_A^{mb}$ with $A = \bigcup \limits^s_{i=1} A_i$, $B=\{(A_{b},B_{b},M_{b}, R_{b})\mid1\leq {b} \leq s_b\}$, $M= \{(A_{m},B_{m},M_{m}, R_{m})\mid s_b+1\leq {m} \leq s\}$ and $R\subseteq J$.
\end{itemize}
\end{defn}

We have incorporated a set of control requirement indices into the Definition \ref{def:2} to streamline the derivation of the MLDES. Each cluster can now include the set of related requirements. The next section describes how this definition is used in the transformation algorithm to the MLDES.

\section{Transformation algorithm}\label{s:algo}
From now on we assume that a multilevel clustering $(I,M,B,J)$ is given on index set $I$ of the most refined product system $\{G_i\mid i\in I\}$. We instantiate the root cluster with the entire index set $J$ of control requirements $\{K_j\mid j\in J\}$, and the children (if any) of the root cluster contain an empty set of requirement indices. The goal is to distribute the control requirements at the relevant nodes, starting with the complete set of control requirements at the root cluster, and decide as we traverse the multilevel clustering.

Algorithm \ref{alg:alg1} is used to build a tree-structure of synthesis subproblems based on the hierarchy defined in multilevel clustering, where the nodes are populated with plant models and requirement models given the DMM. Initially, Algorithm \ref{alg:alg1} is executed by calling TransformCtoT($(I,M,B,J),PR, \{G_i\mid i\in I\}, \{K_j\mid j\in J\}$).
The algorithm performs a preorder traversal of the multilevel clustering, meaning that it first visits the root node, followed by the recursive call for subclusters. 
\begin{algorithm}[t]
    \caption{Transform$C$to$T$ }
    \label{alg:alg1}
    \textbf{Input:} {multilevel clustering $(A,B,M,R)\in C_A^{mb}$, DMM $\mathit{PR}$, and set of most refined product system $G=\{G_i\mid i\in I\}$ and set of controlled requirements $K=\{K_j\mid j \in J\}$. } \\
    \textbf{Output:} {transformation from $C_A^{mb}$ to tree-structure of synthesis subproblems $f: C_A^{mb} \rightarrow 2^G\times2^K$} 
        \begin{algorithmic}[1]
        \IF{size$(A)=1$}
            \STATE $P = A$
            \FOR{$r \in R$}
                \STATE $P = P \cup\{i\in I \mid \mathit{PR}(i,r)=1\}$
            \ENDFOR
            \STATE $f(A,B,M,R)=(\{G_p\mid p\in P\}, \{ K_r\mid r\in R\})$
        \ELSE                  
            \STATE $(A,B,M,R), P =$ PropReq$((A,B,M,R), \mathit{PR}, G)$
            
            \STATE $f(A,B,M,R)=(\{G_p\mid p\in P\}, \{ K_r\mid r\in R\})$
            \FOR{$c \in B\cup M$}  
                \STATE $f(c)$ = Transform$C$to$T$($\mathit{c}, \mathit{PR}, G, K$)         
            \ENDFOR
        \ENDIF
        \end{algorithmic}
    \end{algorithm}
For a leaf cluster, we immediately assign the corresponding plant to the node, together with any associated requirements and the plants to which they refer, as shown in lines 1--7 of Algorithm \ref{alg:alg1}. 
For each non-leaf node in lines 8--11 of Algorithm~\ref{alg:alg1}, Algorithm \ref{alg:alg2} checks for each requirement if it applies to the current cluster or its children. By calling Algorithm \ref{alg:numclusters} for bus and non-bus clusters in line 3 and 4 of Algorithm \ref{alg:alg2}, respectively, we compute related clusters for each requirement. If the requirement mentions the plant that is inside the cluster, then this cluster is related. We push a requirement to a non-bus child cluster if it is the only related non-bus cluster, see lines 6--8 of Algorithm \ref{alg:alg2}. We push a requirement to a bus child cluster if it is the only related cluster, see lines 10--12 of Algorithm \ref{alg:alg2}. Otherwise, a requirement remains in the current cluster, see lines 13--15 of Algorithm \ref{alg:alg2}. After processing all requirements, the remaining requirements along with the plants to which they refer are assigned to the node in line 9 of Algorithm \ref{alg:alg1}. 
    \begin{algorithm}[t]
    \caption{PropReq}
    \label{alg:alg2}
     \textbf{Input:} {multilevel clustering $(A,B,M,R)\in{C_A^{mb}}$, DMM $\mathit{PR}$, set of plant components $G=\{G_i \mid i\in I\}$}\\
     \textbf{Output:} {index set of plants $P$, multilevel clustering $(A,B,M,R)\in{C_A^{mb}}$}
        \begin{algorithmic}[1]
        \STATE Set $ P = \varnothing$
        \FOR{$r\in R$}
            \STATE $\mathcal{B} =$ ComputeRelatedClusters$(B,\mathit{PR},r)$
            
            \STATE $\mathcal{M} =$ ComputeRelatedClusters$(M,\mathit{PR},r)$
            \IF{$|\mathcal{M}| = 1$}
            \STATE Get $(A', B',M', R') \in \mathcal{M}$
            
                \STATE {\small $M= (M\setminus \{(A', B',M', R')\})\cup\{(A', B',M', R'\cup\{r\})\}$}
                \STATE {\small$R=R\setminus\{r\}$}
            \ELSIF{$|\mathcal{B}|= 1 \wedge |\mathcal{M}| = 0$}
            \STATE Get $(A', B',M', R') \in \mathcal{B}$
                \STATE {\small$B= (B\setminus \{(A', B',M', R')\})\cup\{(A', B',M', R'\cup\{r\})\}$}
                \STATE {\small$R=R\setminus\{r\}$}
            \ELSE
                \STATE  $P = P \cup \{i\in I \mid PR(i,r)=1\}$
            \ENDIF
        \ENDFOR     
        \end{algorithmic}
    \end{algorithm}   
\begin{algorithm}[t]
    \caption{ComputeRelatedClusters}
    \label{alg:numclusters}
     \textbf{Input:} {multilevel clustering $C\in C_A^{mb}$, DMM $\mathit{PR}$, requirement $r$}\\
     \textbf{Output:} {set of related cluster $\mathcal{C}$}
        \begin{algorithmic}[1]
        \STATE $\mathcal{C} = \varnothing$
    \FOR{$(A_{i}, B_{i}, M_{i}, R_{i}) \in C$} 
                        \IF{$\exists p\in A_{i}  \text{ s.t. } \mathit{PR}(p,r) = 1$}
                        \STATE $\mathcal{C}:=\mathcal{C}\cup\{(A_{i}, B_{i}, M_{i}, R_{i})\}$
                        \ENDIF
                \ENDFOR
                \end{algorithmic}
\end{algorithm}  
Algorithm \ref{alg:alg1} is called recursively when the cardinality of the current multilevel clustering $(A,B,M,R)$ is larger than 1 for the each bus and non-bus subcluster in line 11. The following proposition guarantees the termination of the algorithm.
\setcounter{thm}{0}
\begin{prop}\label{prop:0}{(Termination of Algorithm \ref{alg:alg1}).}    
Algorithm \ref{alg:alg1} terminates given a valid multilevel clustering $(A,B,M,R)$ with finite sets $A$ and $R$.
\end{prop}
\begin{pf}
Algorithm \ref{alg:alg1} is called inductively when the cardinality of the current multilevel clustering $(A, B, M, R)$ is greater than 1 (line 11). The new inductive calls are performed on subclusters of $B$ and $ M$, where for each subcluster $(A_{i}, B_i, M_i,R_i)$ of $B$ and $M$  it holds that $|A_{i}|<|A|$. We can prove inductively:
 
\quad \textbf{Inductive base:} When $|A| = 1$, the algorithm terminates.

\quad\textbf{Inductive hypothesis:} Assume that $\forall(A_{i}, B_{i}, M_{i}) \in B\cup M$, the algorithm will terminate eventually for $(A_i, B_i, M_i)$.

\quad\textbf{Inductive step:} In Algorithm \ref{alg:alg1} inductive calls are performed $\forall(A_{i}, B_{i}, M_{i}) \in B\cup M$. Given the inductive hypothesis stating that each of these inductive calls will terminate, all inductive calls in Algorithm \ref{alg:alg1} will terminate. Therefore, Algorithm \ref{alg:alg1} with $(A,B,M)$ will terminate.
 \end{pf}
 
Moreover, the algorithm satisfies the following Propositions \ref{prop:1},\ref{prop:2}, and \ref{prop:3}.
\begin{prop}\label{prop:1}{(Valid synthesis problem).} Every non-empty node represents a valid synthesis problem. The synthesis problem is valid if all plant models that are referred in the requirement models are present. In other words, every node that contains one or more requirements also contains all plants involved in those requirements. 
\end{prop}
A proof of Proposition \ref{prop:1} for a node containing only a plant is trivial, and to construct the proof for the nodes containing requirements, we first define the following lemma.
\setcounter{thm}{0}
 \begin{lem}\label{lem:1}{(Valid synthesis problem for non-leaf nodes).}
     Algorithm \ref{alg:alg2} distributes requirements among children nodes (subclusters) or leaves them in the current node. In the latter case, it returns an index set $P$ of plants, which constitute a valid synthesis problem.
 \end{lem}
\begin{pf}
    In Algorithm \ref{alg:alg2}, for each requirement $r$ that remains in the node (lines 13--15), indices of all plants mentioned in the requirements are collected in $P$ (line 14) to be added to the node in Algorithm \ref{alg:alg1} in line 9. In line 9 of Algorithm \ref{alg:alg1} all the requirements are added to the node as well.
\end{pf}
Then we can prove Proposition \ref{prop:1} as follows.
 \begin{pf}
     If we assume that a node contains a requirement, then Algorithm \ref{alg:alg1} always adds involved plants after it inserts requirements. If it is a leaf node, in lines 3-5 of Algorithm \ref{alg:alg1}, all plant components that are involved in the requirements of that node are collected in lines 3-5 and added together in line 6. Otherwise, Algorithm \ref{alg:alg2} is called in line 8 for every non-leaf node. Using Lemma \ref{lem:1}, this algorithm makes sure that the node is populated with the requirements and all the plants involved in them.
 \end{pf}
\setcounter{thm}{2}
\begin{prop}\label{prop:2}{(Plant model conservation).}
All plant models are present somewhere in the output tree at least once.    
\end{prop}

\begin{pf}
    Each leaf node in the input tree represents a single plant. The algorithm assigns that plant directly to the corresponding node in the output tree. So, every plant appears at least in its own leaf node. This can be seen in Algorithm \ref{alg:alg1} in line 2. Additionally, the algorithm never removes plants from the nodes as only the union of plants is used every time a set of plants is updated (see Algorithm 1 (line 4) and Algorithm \ref{alg:alg2} (line 14)).
\end{pf}

\begin{prop}\label{prop:3}{(Requirement model conservation).}
All requirements models are present somewhere in the output tree once.
Invariant, that is preserved each step: \\$\dot {\bigcup\limits_{n\in Nodes((A,B,M,R))}} n.R = R$, where:
    \begin{itemize}
        \item $R$ is the index set of all requirements for the system that need to be distributed among bus and non-bus clusters.
        \item $n.R$ is the index set of requirements in a current node $n$.
        \item $Nodes((A,B,M,R))=\\ \phantom{space}\{(A,B,M,R)\}\cup \bigcup\limits_{b\in B}Nodes(b)\cup \bigcup\limits_{m\in M}Nodes(m)$
    \end{itemize}    
\end{prop}

\begin{pf}
    Initially, the invariant holds, because in multilevel clustering $(A,B,M,R)$ $R$ contains all the requirements, while $\forall i (A_{i} , B_{i} , M_{i} , R_{i} ) \in B, R_{i}=\varnothing$ and  $\forall j (A_{j} , B_{j} , M_{j} , R_{j} ) \in M, R_{j}=\varnothing$.  If a node $(A,B,M,R)$ has children, Algorithm \ref{alg:alg2} is called, where requirements $R$ are distributed among the node and its children in two cases:
\begin{enumerate}
    \item requirement $r$ involves plant components that are in $A'$ of one child node ($A', B',M',R'$) of $M$, then this requirement $r$ is moved from $R$ to $R'$ (see lines 6-8)
    \item requirement $r$ involves \textbf{only} plant components that are clustered in one child node ($A', B',M',R'$) of $B$, then this requirement $r$ is moved from $R$ to $R'$ (see lines 10-12)
\end{enumerate}
Otherwise, requirement $r$ remains in $R$ (see line 11). Algorithm \ref{alg:alg2} moves requirements from parent's $R$ to its children's $R'$ while traversing following preoder traversal strategy, and no requirement is added or lost. First, the root cluster is visited, followed by each bus and non-bus subcluster, so requirements are being distributed until the leaf node is reached which owns the remaining requirements.
\end{pf}


\section{Case study}\label{s:example}
As a case study, a small-scale production line model is considered. 
\begin{figure}[h]
    \centering
    \includegraphics[width=\linewidth]{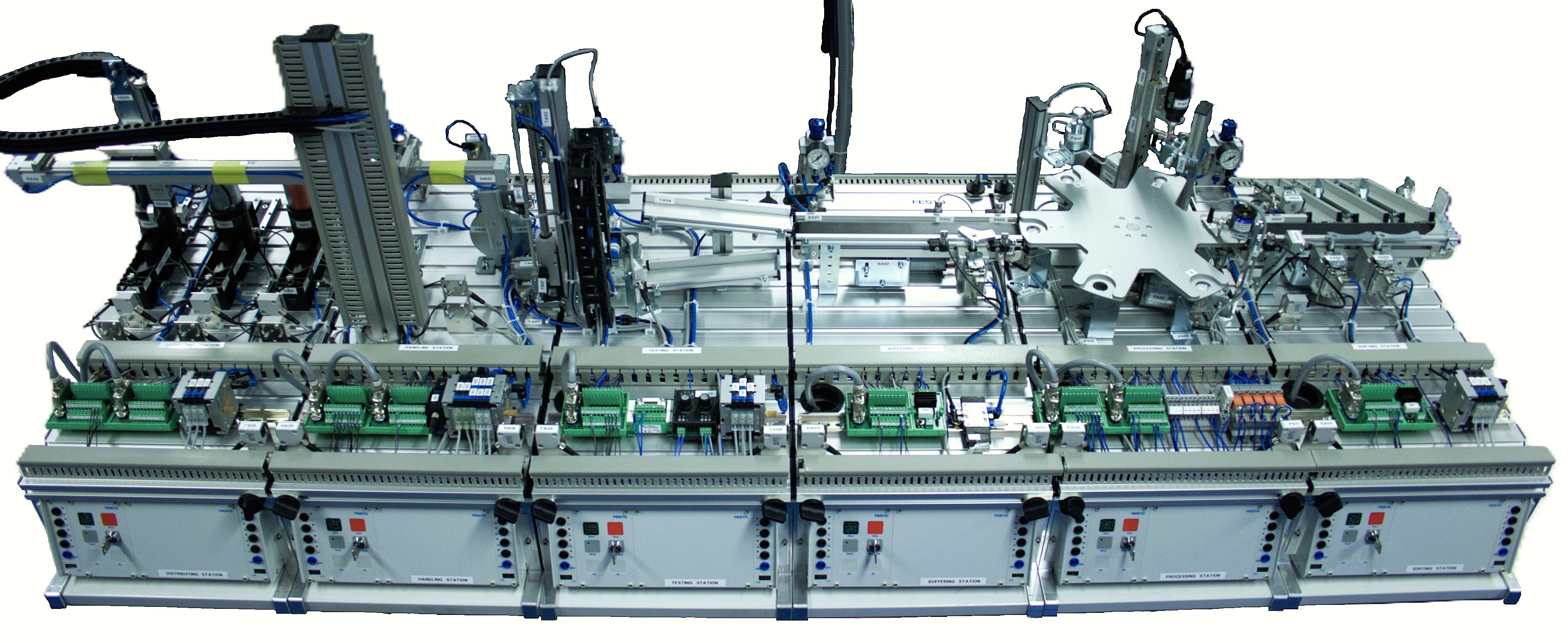}
    \caption{Festo station.}
    \label{fig:festo}
\end{figure} Fig.~\ref{fig:festo} illustrates the production line, consisting of six stations (ordered from left to right):
\begin{enumerate}
    \item \textbf{Distributing Station} -- Products are initially placed in three tubes and released via pushers.
    \item \textbf{Handling Station} -- A pneumatic gripper transports products to an intermediate buffer.

    \item \textbf{Testing Station} -- Product height is measured; correct products move on, rejects go to a local buffer.

    \item \textbf{Buffering Station} -- Conveyor belt and separator manage the product flow towards processing.
    
    \item \textbf{Processing Station} -- A turntable with six positions checks orientation, performs (virtual) drilling, and directs products further.
    
    \item \textbf{Sorting Station} -- As the final step, products are classified by color and material, then diverted into one of three buffers.
\end{enumerate}

This small-scale production line is part of Festo Didactic (\url{www.festo.com}) learning systems for mechatronics training, namely the Modular Production System (MPS). Although it does not carry out real processing, each station is equipped with industrial actuators and sensors reproducing the dynamics of industrial-related processes. For actuation, various types of pneumatic cylinders and DC motors are employed, with a total of 28 actuators integrated into the stations. For sensing -- optical, inductive, capacitive, electromechanical, reed contact, and Hall-effect sensors, with total of 59 sensors.

\begin{figure}[h]
        \centering
        \includegraphics[width=\linewidth]{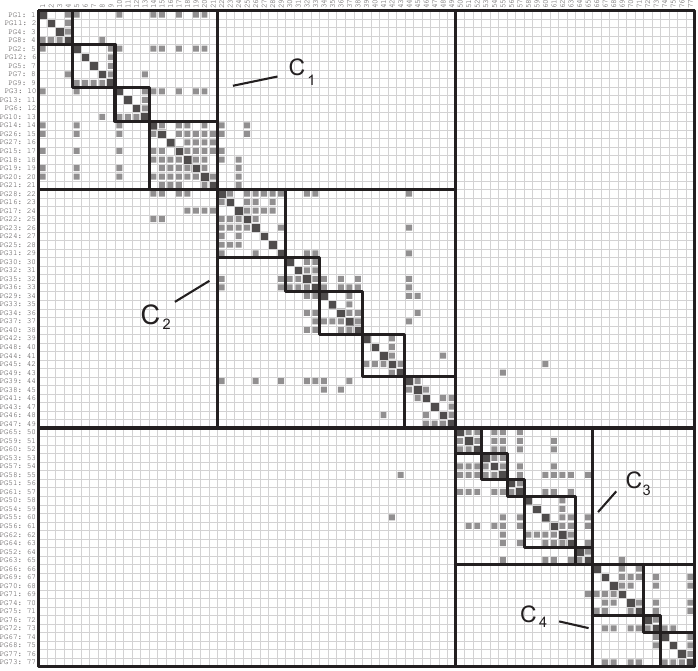}
        \caption{The result of DSM-based clustering of the Festo production line without bus. On the third-level, clusters $C_1$, $C_2$, $C_3$, and $C_4$ are marked.}
        \label{fig:dsm1}
\end{figure}
\begin{figure}[h]
        \centering
        \includegraphics[trim = 0mm 37mm 0mm 36mm, clip, width=\linewidth]{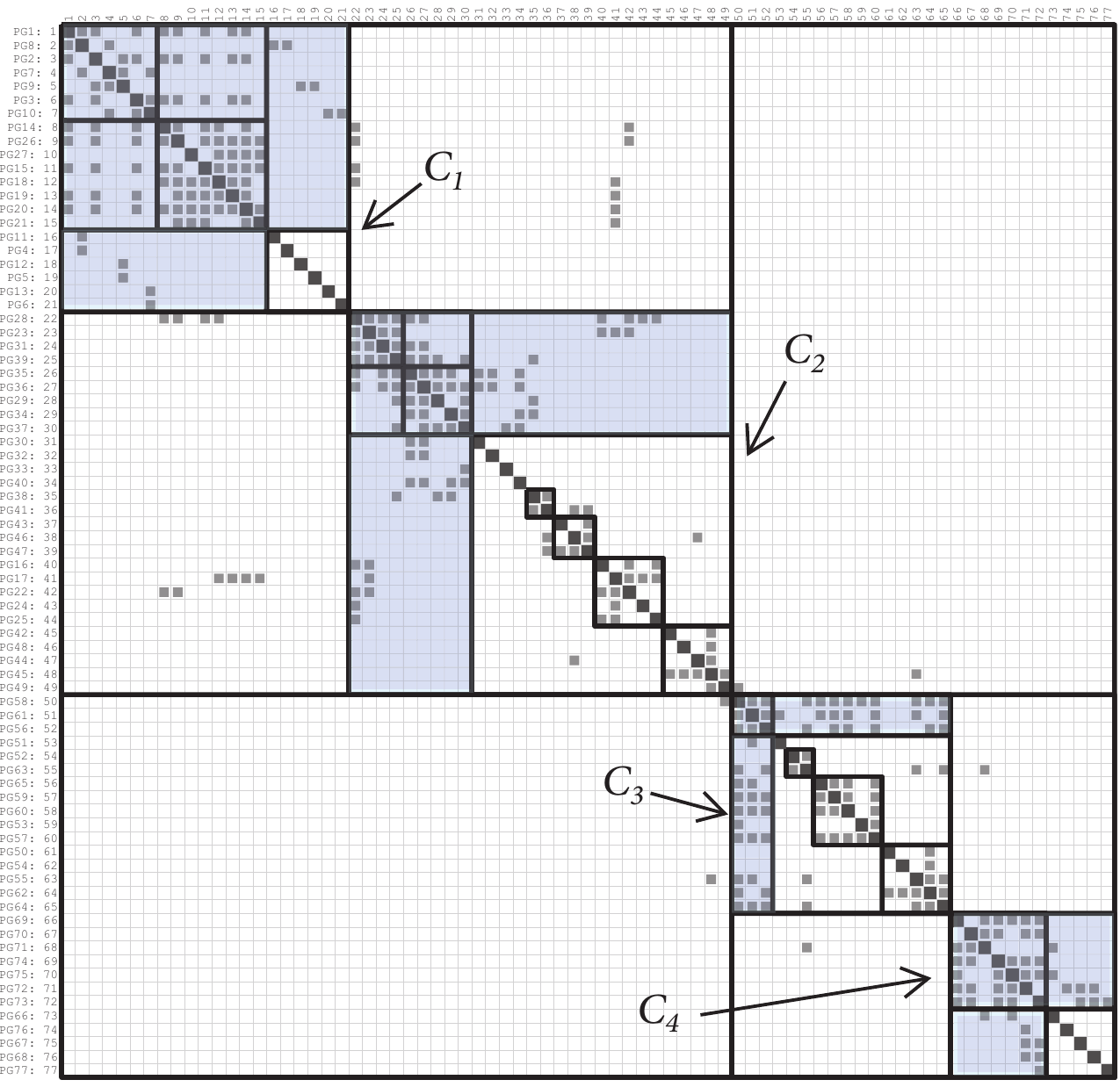}
        \caption{The result of DSM-based clustering of the Festo with local buses. Bus detection is applied on clusters $C_1$, $C_2$, $C_3$, and $C_4$. The local bus for each cluster is highlighted in blue.}
        \label{fig:dsm2}
\end{figure}

\cite{Reijnen-ccta18} performed a DSM-based clustering of the production line, which resulted in four clusters found on the third level labeled $C_1$, $C_2$, $C_3$ and $C_4$ (see Fig.~\ref{fig:dsm1}). The distribution station with some components of the handling station with which it interacts is clustered in $C_1$. $C_2$ includes the buffering and testing stations with some components of the handling and processing stations. The clusters $C_3$ and $C_4$ contain components of processing and sorting stations, respectively. Obtaining MLDES from this architecture results in 101 synthesis subproblems. After subtracting single-plant specifications that have already been mentioned, the MLDES would require 24 supervisors, with which the sum of the controlled state-space size of the multilevel supervisors equals $50,638$ compared to $2.2\times 10^{25}$ in monolithic synthesis. Furthermore, \cite{Goorden-ecc19} has experimentally shown that this multilevel clustering yields better multilevel synthesis performance compared to MLDES with the global bus with controlled state-space size of $2.6\times 10^{5}$. 

To further reduce this number, the proposed method is applied. The clusters $C_1$, $C_2$, $C_3$, and $C_4$ have been re-evaluated with local bus detection clustering using the same tuning parameters. The resulted architecture is depicted in Fig.~\ref{fig:dsm2}. The new local bus MLDES would require 50 supervisors with the sum of the controlled state-space size reduced to $7,393$.

\begin{figure}[h]
    \centering
    \includegraphics[width=\linewidth]{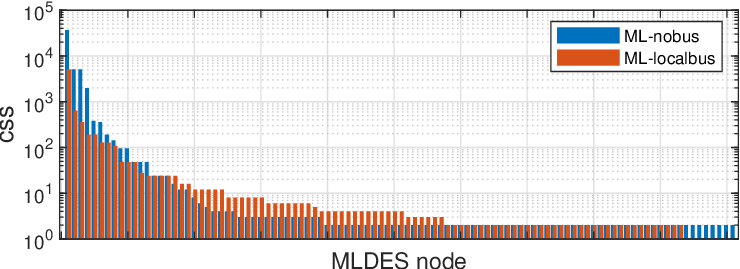}
    \caption{Controlled state-space size of the multilevel supervisors in two MLDESs with and without local buses.}
    \label{fig:distribution}
\end{figure}

In Fig.~\ref{fig:distribution} the difference in the controlled state-space size of multilevel supervisors (sorted in descending order across x-axes) obtained from these two architectures is shown. Although introduction of the notion of local buses to MLDES entails a modest increase in the computational effort of lower-level subproblems, it yields a more balanced decomposition. The total computational effort is reduced, and the distribution of effort across the suggested structure smooths the peak. 

\section{Concluding remarks}\label{s:conclusion}
The paper offers a new concept in the construction of MLDES with local buses to further reduce the computational effort of synthesizing supervisors. 
Future research can focus on improving the performance by discovering the optimal multilevel clustering.
The multilevel clustering of the system was shown to influence the decomposition of the synthesis problem. In our case study, the same clustering parameters were applied to every third-level subcluster to identify local buses. It would be interesting to investigate whether using individual clustering parameters for each subcluster could further enhance performance, and to develop a heuristic approach for determining the parameter values that yield the best results. 


\begin{ack}
The authors thank Johan van den Bogaard, Yigal Levin, and Wendy Clerx from RWS for supporting this research.
\end{ack}


\bibliography{ifacconf}             

\end{document}